\documentclass[12pt,preprint]{aastex}

\newcommand{\gtappeq}{\raisebox{-0.6ex}{$\,\stackrel
{\raisebox{-.2ex}{$\textstyle >$}}{\sim}\,$}}

\usepackage{psfig}

\slugcomment{accepted for publication in ApJ Letters}

\shorttitle{Discovery of 15-s~oscillations in WZ~Sge}
\shortauthors{Knigge et al.}

\begin{document}



\title{Discovery of 15-second oscillations in {\em Hubble Space Telescope} 
observations of WZ Sagittae following the 2001 outburst$^1$}


\author{C. Knigge$^2$, R. I. Hynes$^2$, D. Steeghs$^2$,
K. S. Long$^3$, S. Araujo-Betancor$^2$, T. R. Marsh$^2$} 

\email{christian@astro.soton.ac.uk, rih@astro.soton.ac.uk,
ds@astro.soton.ac.uk, long@stsci.edu, sab@astro.soton.ac.uk,
trm@astro.soton.ac.uk}

\begin{abstract}

We report the discovery of 15-s oscillations in ultraviolet
observations of WZ~Sge obtained with the {\em Hubble Space Telescope}
approximately one month after the peak of the 2001 outburst. This is
the earliest detection of oscillations in WZ~Sge following an outburst
and the first time that a signal near 15~s has been seen to be
dominant. The oscillations are quite strong (amplitude $\simeq 5$\%),
but not particularly coherent. In one instance, the oscillation
period changed by 0.7~s between successive observations separated by
less than 1~hour. We have also found evidence for weaker signals with
periods near 6.5~s in some of our data. We discuss the implications of 
our results for the models that have been proposed to account for the
28-s oscillations seen in quiescence. If the periods of the 15-s
oscillations can be identified with the periods of revolution of
material rotating about the white dwarf, the mass of the white dwarf
must satisfy $M_{WD} > 0.71~M_{\odot}$. The corresponding limit for
the 6.5-s signals is $M_{WD} > 1.03~M_{\odot}$. 

\end{abstract}


\footnotetext[1]{Based on observations with the NASA/ESA Hubble Space
Telescope, obtained at the Space Telescope Science Institute, which is
operated by the Association of Universities for Research in Astronomy,
Inc. under NASA contract No. NAS5-26555.}

\footnotetext[2]{Department of Physics \& Astronomy, University of
Southampton, Highfield, Southampton SO17 1BJ, UK} 

\footnotetext[3]{Space Telescope Science Institute, 3700 San Martin
Drive, Baltimore, MD 21218, USA}

\section{Introduction}

The dwarf nova WZ Sge is arguably the most extreme cataclysmic
variable (CV) known. Whereas other dwarf novae undergo 3-5
magnitude eruptions every few weeks or months, WZ 
Sge's outbursts have an amplitude of 7-8 magnitudes and recur on a
time-scale of roughly 33 years. WZ Sge's 82 minute
orbital period is also one of the shortest of any CV, its mass ratio is 
one of the lowest ($q =M_2/M_{WD} \simeq 0.05$; Steeghs et al. 2001) 
and its time-averaged absolute magnitude one of the faintest ($M_V
\simeq 11.5$; Patterson 1998). All of these facts suggest that WZ Sge
is a highly evolved CV whose secondary is probably a brown dwarf-like
object (Patterson 1998). 

Many studies of WZ~Sge in quiescence have found oscillations near
28~s (Robinson, Nather \& Patterson 1978; Patterson 1980;
Skidmore et al. 1997; Welsh et al. 1997; Patterson et al. 1998 [P98];
Skidmore et al. 1999 [S99]). However, the origin of 
these oscillations is still unclear: Robinson
et al. (1978) and S99 favour a ZZ Ceti-like, pulsating white dwarf
(WD) model, but Patterson (1980) and P98 prefer an oblique
magnetic rotator model. A modified version of the latter model has
recently been proposed by Warner \& Woudt (2002 [W02];
also see Section~\ref{discuss}). 

Here, we report the discovery of 15-s oscillations in HST
observations of WZ~Sge obtained roughly 1 month after the start of its
2001 outburst. This is the earliest detection of rapid oscillations
following an outburst of the system (Patterson et al. 1981).

\section{Observations}

WZ~Sge went into outburst on July 23, 2001 (Ishioka et
al. 2001). The ensuing multi-wavelength campaign included two {\em
Hubble Space Telescope} (HST) {\em Director's Discretionary Time}
(DDT) programs. The HST observations analysed here were obtained with
the {\em Space Telescope Imaging Spectrograph} (STIS) during the first
DDT program. The goal of this program was to cover the immediate
aftermath of the outburst in the FUV, with three observing epochs on
Aug 8, Aug 19 and Aug 22. Each epoch consisted of 4 consecutive HST
orbits. The bulk of
the observing time in each epoch was used to obtain time-resolved,
high-resolution, FUV  
spectroscopy of WZ~Sge with the E140M echelle grating dispersing the
light onto the FUV-MAMA detectors operating in TIME-TAG mode.
This set-up covers the region
1150~\AA~--~1700~\AA~at a spectral resolution of
45,800. Due to overheads, FUV exposure times in the first and third
orbit of each epoch are shorter than those in the second and fourth
orbits. 

Figure~1 illustrates the timing of the HST observations
relative to the optical outburst light curve. Following the
nomenclature of Patterson et al. (2002), we see that the first of our
HST observations occurred during the plateau phase of the outburst,
the second during the ensuing dip and the third near the peak of 
the first echo outburst.

\section{Analysis}

Since our focus here is on the search for rapid oscillations, we 
constructed ``white light'' light curves at 1-s time resolution
directly from the TIME-TAG files. These files contain a
list of the arrival times and detector positions of all recorded photon
events. The backgrounds due to dark current and geocoronal
emission are negligible in our data, so the resulting light curves 
are ideally suited for studying the short-timescale variability of
WZ~Sge.

The raw FUV light curves immediately revealed the presence of strong,
rapid oscillations in Epoch~3, though not in Epochs~1 and 2. The
Epoch~3 light curves are shown in 
Figure~2. A roughly 5\% oscillation with period near 15~s is
easily visible. The oscillation amplitude is clearly variable (e.g. note
the weakness of the oscillations around $t = 700$~s and $t
=12,400$~s). However, there 
is no obvious link between oscillation amplitude and orbital
phase $\phi_{orb}$ (computed from the ephemeris of P98; inferior
conjunction of the secondary corresponds to $\phi_{orb} \simeq -0.046$ 
[Steeghs et al. 2001]).

Figure~3 shows the discrete Fourier transform of each
HST orbit within Epoch~3. As expected, signals near 15~s are easily
seen in all data subsets. No obvious signals are seen at frequencies
corresponding to periods around 30~s. In particular, there is no
convincing evidence for the 27.87-s signal that is the most stable
clock in quiescence (P98), nor for any ``subharmonics'' of the dominant
15-s signals. Figure~3 also reveals that the
oscillation period is variable. Most noticably, the dominant period in 
Orbit~1 is $P_1 = 13.93$~s, but this has changed to $P_2 = 14.67$~s in
Orbit~2.

Signals with periods of about 6.5~s are also present in Orbits 1
and 3. If these are harmonics of signals near the dominant 15-s
periods, then at least in the case of  the 6.43-s signal in Orbit~1,
the harmonic has to dominate 
strongly over the fundamental. The 6.5-s oscillations appear to
be strongest when the 15-s oscillations are weakest. Thus in the raw
light curves (Figure~2), the 6.5-s oscillations are only
discernible when the main signal is weak, e.g. near $t=700$~s and
$t=12,400$~s.

In order to analyse the time evolution of the 15-s oscillations more
carefully, we carried out sliding sinusoid 
fits to successive chunks of the Epoch~3 light curve. The model 
was a constant period sinusoid superposed on a constant DC offset,  
$y(t) = a \cos{(2 \pi f t - \phi)} + {\rm DC}$.
In practice, $f$ was always fixed to the dominant frequency in each 
orbit (see Figure~3). The fits were 
carried out to successive 45-s chunks of data, with no overlap. The
phase of the first chunk in each orbit was arbitrarily set to
180$^{\circ}$.

The results are shown in Figure~4. There
appears to be no connection between the amplitude of the oscillations
and the average flux (the DC offset). The median oscillation amplitude
is 530~c/s; the median ratio of the amplitude to the DC
offset is 5.0\%. The obvious phase variability in
Figure~3 confirms that the
oscillation period is variable. More specifically,
given the form of our model, a linear phase shift implies that the
true period differs from the trial period. 
Any non-linearity in the phase plots is a sign that the 
oscillation period is changing. For example, a transition from one
linear portion to another indicates a switch from one period to 
another. The phase plots in Figure~4 clearly show such 
transitions, usually on time-scales shorter than we can resolve. We
emphasise, however, that our fits assume a {\em single}, sinusoidal
signal. If {\em several} signals are present simultaneously, the
results of such a fit will depend on the relative amplitudes of the
signals and the closeness of their periods to that assumed in the
fit.

\section{Discussion}

\label{discuss}

We begin our discussion by summarising the key properties of the
15-s oscillations we have discovered:
\begin{enumerate}
\item[(i)] they were found in data obtained 1~month after the start
of WZ~Sge's 2001 outburst, just past the peak of the first echo
outburst; 
\item[(ii)] they are strong, with fractional amplitudes around 5\%;
\item[(iii)] their periods are about a factor of two shorter than those 
of the 28-s oscillations seen in quiescence; the latter are not
seen in our data;
\item[(iv)] they exhibit phase jitter consistent with small, possibly
discontinuous period changes.  
\end{enumerate}
In addition, we have discovered 6.5-s oscillations in two of our four
HST orbits.

These properties are unusual. Oscillations in WZ~Sge have previously
only been seen in data obtained in quiescence. Their periods are
typically 28~s -- 29~s (S99), with the most stable signal lying at 28.87~s 
(P98). Oscillations near 15~s have only been seen once
before (Provencal \& Nather 1995), but never as the dominant
signal. Finally, no oscillation with $P \simeq 6.5$~s has ever been
reported in  WZ~Sge.\footnote[4]{We note in passing that Eracleous, Patterson \&   
Halpern (1991) found marginal evidence for a 9.56-s periodicity in
Einstein x-ray data.}.

It is interesting that the dominant periods in our data are roughly
half of those seen in quiescence. However, only the period determined for 
Orbit~1 could be harmonically related with the 27.87~s signal ($2
\times 13.93$~s~$= 27.86$~s). Among other periods reported in the literature 
(see Table~2 in S99), the closest harmonic matches to the periods
in our data are: 29.33~s (Orbit 2: $2 \times 14.67$~s~$= 29.34$~s); 
29.10~s  (Orbit 3: $2 \times 14.60$~s~$= 29.20$~s); 29.69~s (Orbit 4:
$2 \times 14.78$~s~$= 29.56$~s). However, $\sim$~10
apparently distinct periods around 28~s/29~s have been reported, so
the statistical significance of these near matches is unclear. Figure~3
certainly reveals no evidence of any $\simeq$ 29-s fundamental periods
in our data. Moreover, if the peaks in Figure~3 near 6.5~s are also
interpreted as harmonics, the corresponding fundamentals (Orbit 1: $4
\times 6.43$~s~$= 25.72$~s; Orbit 3: $4 \times 6.64$~s~$= 26.56$~s)
have no counterparts among previously reported periods. 

It seems unlikely that the oscillations in our data can be due to
ZZ~Ceti-like WD pulsations. A preliminary analysis of the mean FUV
spectrum obtained during Epoch~2 (only 3 days before Epoch~3),
suggests $T_{WD} \simeq 
25,000$~K (Kuulkers et al. 2001; K.S. Long 2002, private communication).
This is well beyond the blue edge of the ZZ~Ceti instability strip at
$T_{WD} \simeq 13,500$~K (for $M_{WD} \simeq 0.8~M_{\odot}$; Bradley
\& Winget 1994). However, we cannot rule out the pulsation
model definitively, because the response of a pulsating WD to the 
heating and compression it experiences during a dwarf nova outburst is
unknown.

In order to examine the magnetic rotator model, we need some
constraints on the accretion rate through the disk  
in outburst, relative to the quiescent value. Patterson et al. (2002) 
find that the mass transfer rate {\em from the secondary} in
quiescence is $\dot{M}_{2} \sim 10^{15}$~g/s. The quiescent rate must
therefore satisfy $\dot{M}_{q} << 10^{15}$~g/s, and we will adopt
$\dot{M}_{q} < 2 
\times 10^{14}$~g/s as a conservative upper limit. The quiescent x-ray 
luminosity of WZ~Sge is $L_{x} \simeq 
10^{30}$~ergs/s (Eracleous, Halpern \& Patterson 1991; Mukai \&
Shiokawa 1993).\footnote[5]{The quoted
value for $L_x$ has been adjusted to reflect the recent parallax-based 
distance estimate of $d \simeq 45$~pc (J.R. Thorstensen 2002, private
communication)} 
Since the x-rays probably arise
in the boundary layer, they can represent at most half of 
the total quiescent accretion power, i.e. $L_x \leq
(GM_{WD}\dot{M}_{q})/(2R_{WD})$. Using the Nauenberg (1972)
approximation to the Hamada-Salpeter (1961) WD mass-radius, this
yields a lower limit $\dot{M}_{q} \gtappeq 2 \times 10^{12}$~g/s. 
By contrast, a preliminary disk model fit to the time-averaged FUV
spectrum obtained in our Epoch~3 suggests $\dot{M}_{E3} \simeq 3
\times 10^{16}$~g/s (K.S. Long 2002, private communication). Thus the
accretion rate in Epoch~3 exceeded the quiescent rate by a factor in
the range 150 -- 15,000.

This immediately rules out a standard intermediate polar
(IP) model for the oscillations. This is because for a WD satisfying 
$M_{WD} > 0.7~M_{\odot}$ (Steeghs et al. 2001), the 27.87-s
oscillations correspond to 
Keplerian periods at $R < 2 R_{WD}$. This presumably 
marks the inner edge of the quiescent disk in the IP model. However,
the magnetospheric radius of an accreting magnetic WD 
scales as $R_{m} \propto \dot{M}^{-\frac{2}{7}}$ (dipole) and $R_{m}
\propto \dot{M}^{-\frac{2}{11}}$ (quadrupole) (e.g. W02). With
$\dot{M}_{E3}/\dot{M}_{q} > 150$, the magnetosphere 
should thus have been completely crushed onto the WD surface during
Epoch~3.

W02 have recently proposed an alternative magnetic rotator
model. They envisage the magnetic field being anchored in an 
equatorial accretion belt surrounding the WD. This belt has a
relatively low moment of inertia, so its angular velocity can respond
quickly to changes in $\dot{M}$. Their model
predicts a much weaker dependence of $R_m$~on $\dot{M}$, because the 
belt's magnetic field scales as $B \propto \Omega^{1/2} M_b^{1/4}$,
where $\Omega$ is the angular velocity and $M_b$ the mass of the 
belt (W02). According to W02, the resulting dependence of
oscillation period $P$ on $\dot{M}$ can be as weak as $P
\propto \dot{M}^{-1/10}$ during an outburst. Thus the
outburst/quiescence period ratio of about 0.5 corresponds to a ratio
of accretion rates of roughly 1000 in their model. This is within the
allowed range.
\footnote[6]{This argument assumes that the roughly 2:1 period ratio
reflects a change in the {\em fundamental} oscillation period. If it
reflects a switch from the fundamental to the  
first harmonic, the fundamental must be almost unchanged
between outburst and quiescence. Given the huge change in $\dot{M}$
between these states, this would seem hard to explain in any model in
which the oscillation period is partly determined by the ram 
pressure of the accreting material.}

W02 account differently for what they call ``period
discontinuities'', i.e. relatively small period changes on short
time-scales. They propose that the accretion belt is rotating
differentially, with $\Omega$~decreasing away from the
equator. The instantaneous oscillation period is then determined by 
which part of the belt is currently being ``fed''. Period
discontinuities occur as magnetic reconnection events switch the
feeding from one belt region to another. The upshot is that only
longer-term period changes reflect changes in
$\dot{M}$. In WZ~Sge, the change in period
between quiescence and outburst would then probably reflect a change
in $\dot{M}$, but the smaller changes within Epoch~3 would be caused
by magnetic reconnection events on the differentially rotating
accretion belt. 

The ability of the W02 model to account for both the long-term and
short-term period changes is encouraging. However, the model owes
its success at least partly to the sheer complexity of the
mechanism(s) it uses to explain different observational
features. Also, we have not even attempted to explain the 6.5-s
signals in some of our data, nor the weakness/absence of the 
oscillations during Epochs~1 and 2.\footnote[7]{We have not yet
carried out an exhaustive analysis of the Epoch~1 and 2 observations,
but it is clear that there are no signals of comparable strength in
these data sets.}
We therefore feel that the 
origin of the WZ~Sge's oscillations remains an open question.

We finally note that the oscillations can be used to set a lower limit
on the WD mass, under the assumption that their periods they can be
identified with the periods of revolution of material rotating about
the white dwarf. The shortest possible Keplerian period around a  
WD of mass $M_{WD}$ and radius $R_{WD}$ is 
\begin{equation}
P_{K,min} =\frac{2 \pi R_{WD}}{(G M_{WD}/R_{WD})^{1/2}}.
\end{equation}
Again using the Nauenberg (1972) mass-radius relation for cold
WDs, we find a lower limit 
of $M_{WD} > 0.71~M_{\odot}$ for our shortest dominant period of 
13.93~s and $M_{WD} > 1.03~M_{\odot}$ for the weaker 6.43-s signal
seen in Orbit 1. Both limits are consistent with the
constraint $M_{WD} > 0.70 M_{\odot}$ imposed by the spectroscopic mass 
function (Steeghs et al. 2001).

\acknowledgments

This work was supported by NASA through grant G0-9287 from the Space
Telescope Science Institute, which is operated by AURA, Inc., under
NASA contract NAS5-26555. DS and SAB are supported by a PPARC
Fellowship and Studentship, respectively. RIH is supported by Grant
F/00-180/A from the Leverhulme Trust. We are grateful to the VSNET 
collaboration for providing us with the data shown in Figure~1. We
would finally like to thank the many colleagues and amateur observers
who have contributed to the 2001/2002 campaign on WZ~Sge.


\clearpage

\figcaption[f1]{The VSNET optical light curve of the 2001 outburst of
WZ~Sge (see Ishioka et al. 2002). The three HST observing epochs are
marked by arrows.} 

\figcaption[f2]{The FUV light curves for the four HST orbits in
Epoch~3. Note that Orbits 2 and 4 yielded longer time series than
Orbits 1 and 3, so the x-axis for these orbits spans a longer time
interval.}  

\figcaption[f4]{Discrete Fourier transforms of the individual orbits in
Epoch~3. Ordinates correspond to the amplitudes of pure
sinusoids. The dominant frequencies near 15~s and 6.5~s (if
present) are marked. The arrows show the
expected locations of the ``subharmonics'' of these signals. The
dotted line marks the frequency of the 28.87-s signal often seen in
quiescence.}

\figcaption[f5]{The results of the sliding sinusoid fits to the Epoch~3
data. Each point represent the result of a sinusoid fit to a
45-s data chunk; there is no overlap between successive chunks.
The periods of the sine waves were always fixed at the dominant period
of the corresponding orbit (see Figure~3). The top
panels shows the DC offset levels (effectively the mean count rate
within each chunk). The middle panels
show the oscillation amplitudes, and the bottom panels show the
oscillation phases (see text for details). The error bars correspond
to the formal 1-$\sigma$~errors returned by the least squares fits. }

\clearpage

\begin{figure}[t]
\plotone{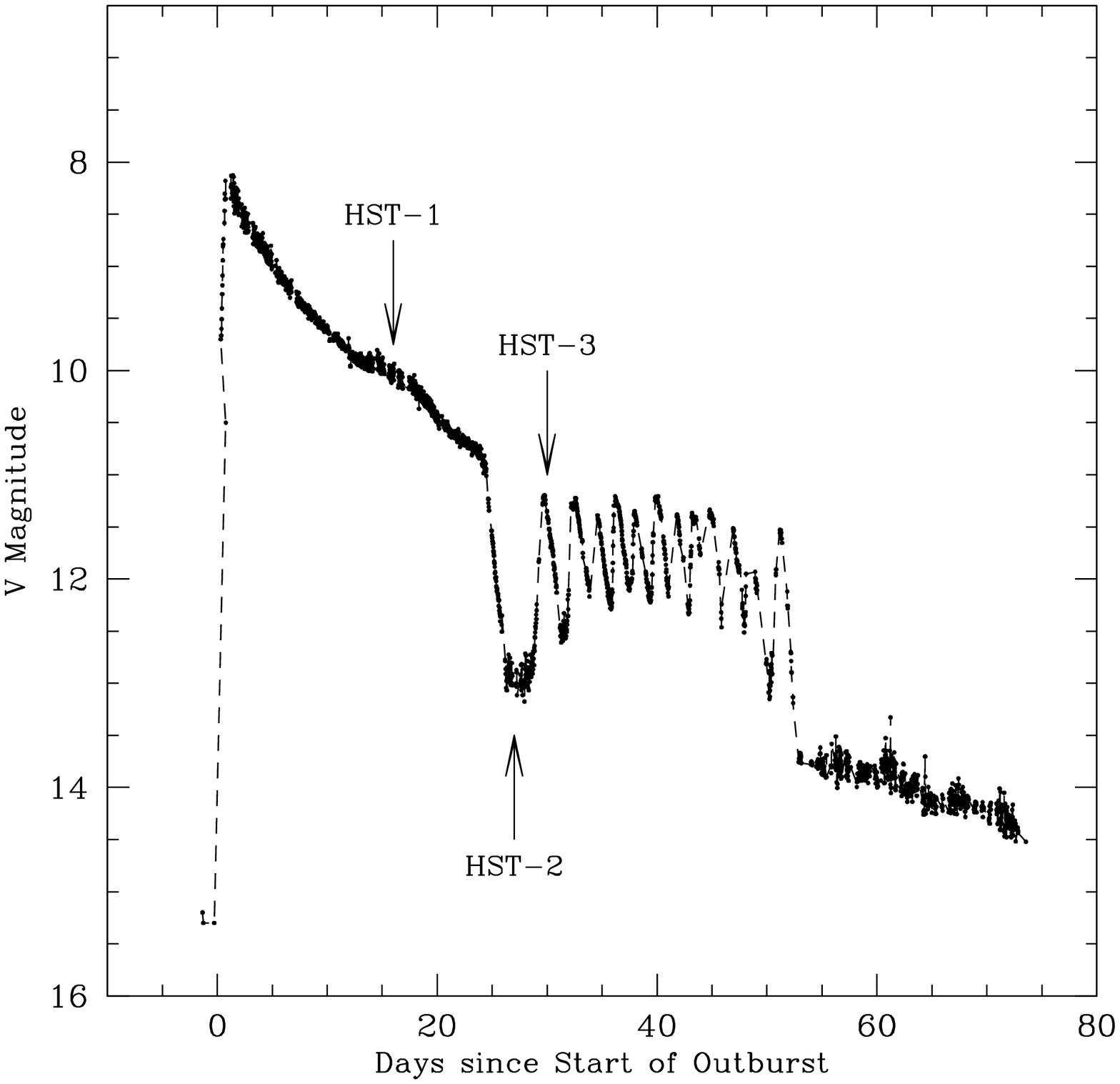}
\label{longlight}
\end{figure}

\clearpage

\begin{figure}[t]
\plotone{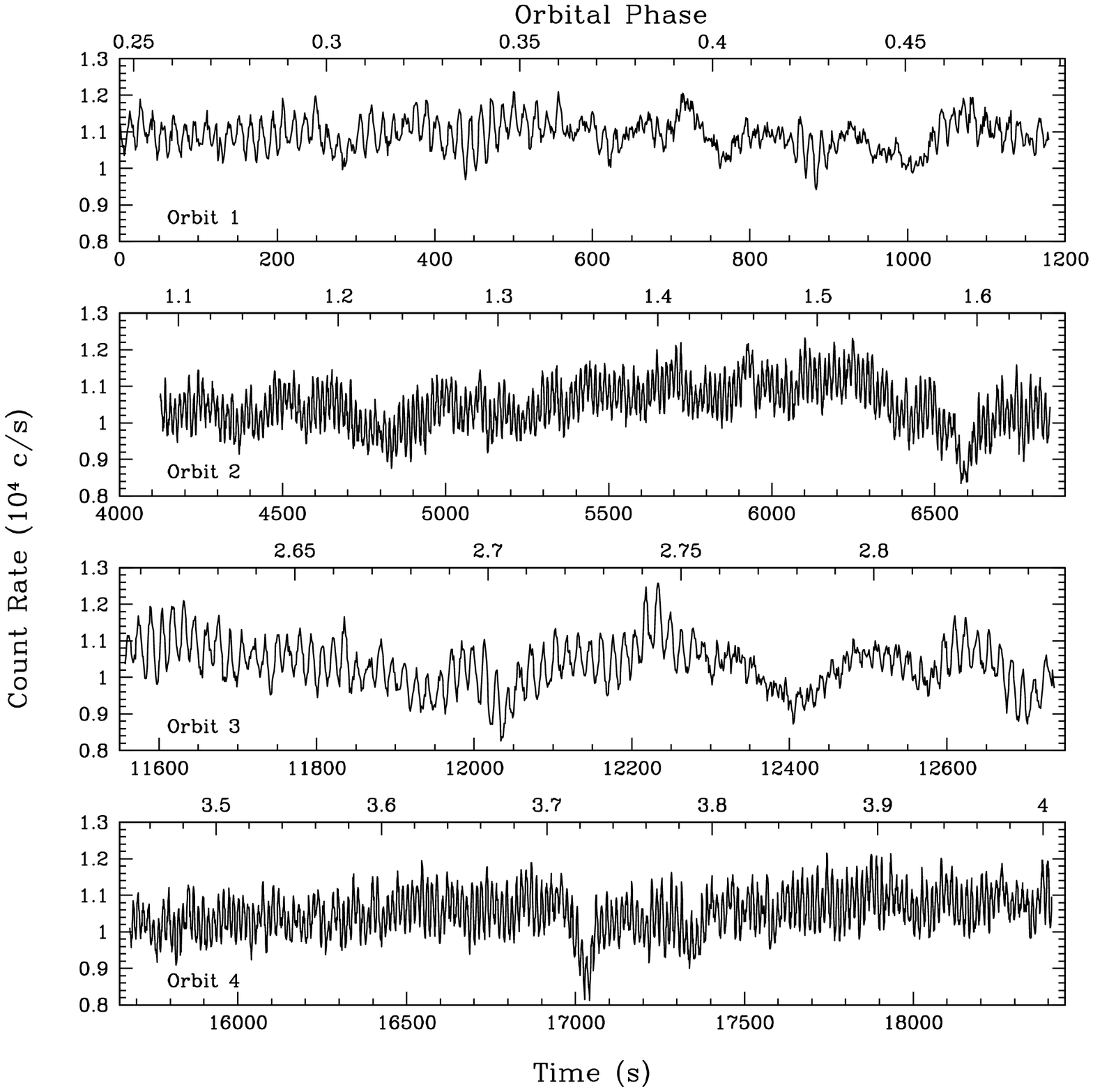}
\label{light}
\end{figure}

\clearpage

\begin{figure}[hb]
\plotone{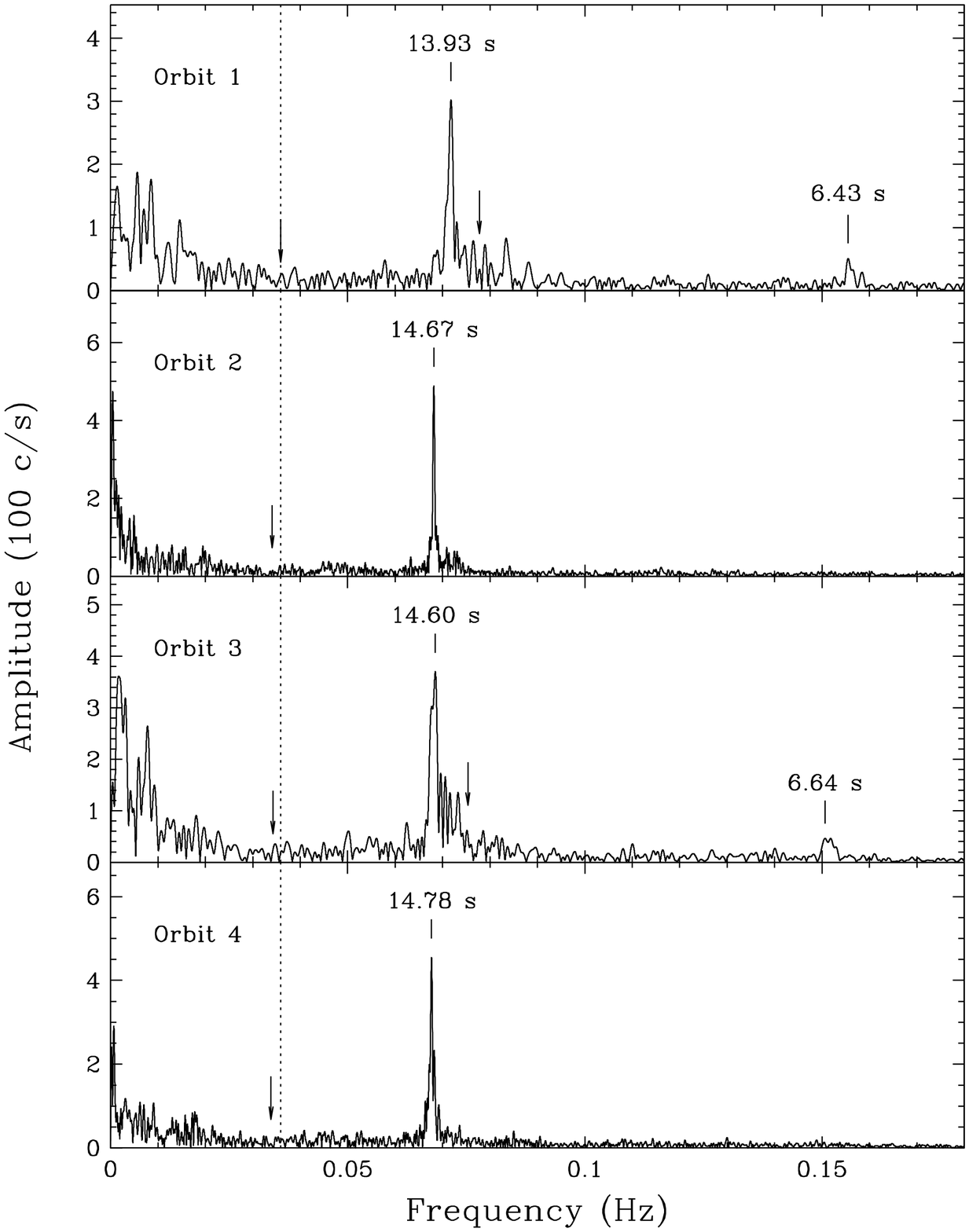}
\label{orbitpower}
\end{figure}

\clearpage

\begin{figure}[t]
\plotone{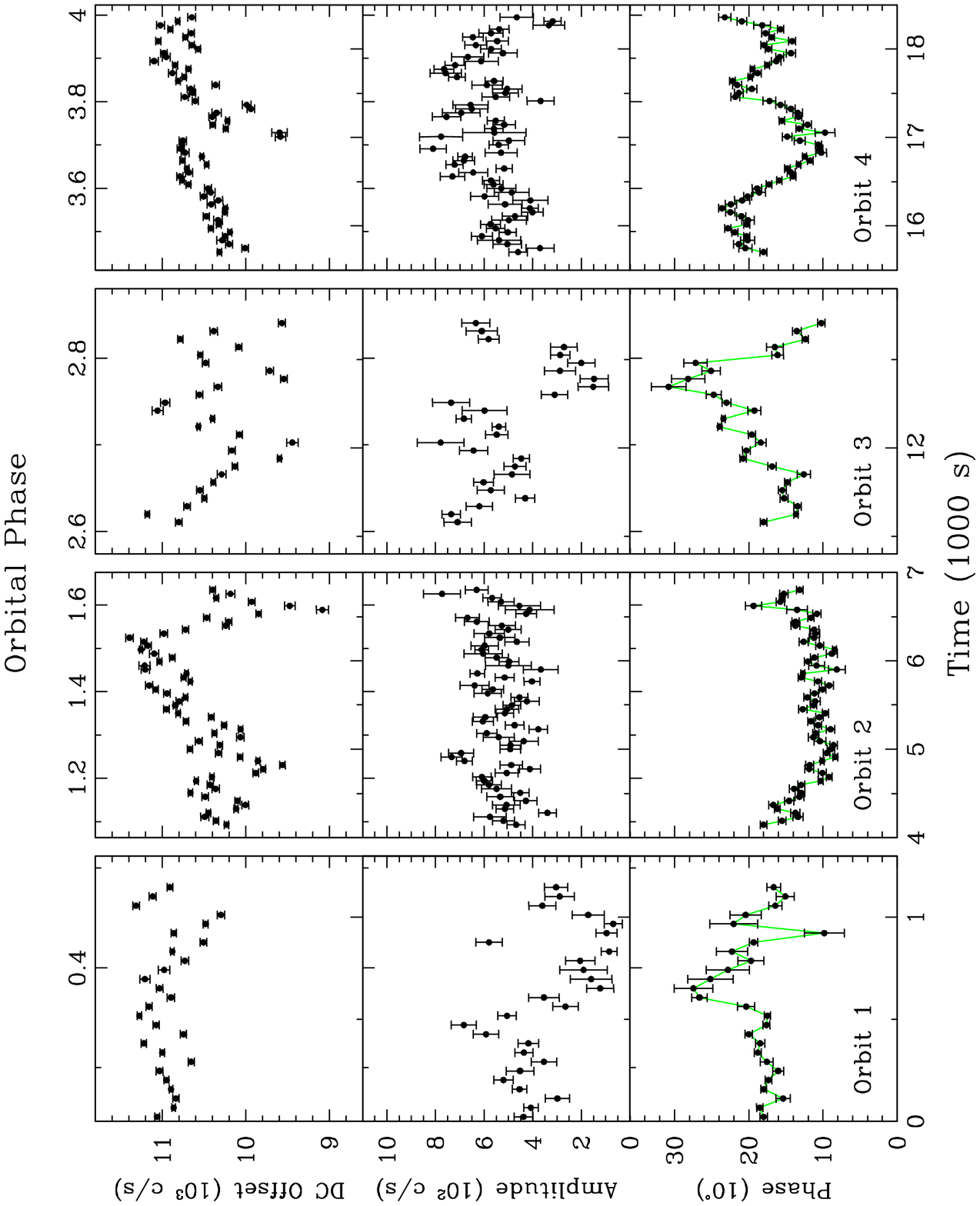}
\label{slidesine}
\end{figure}


\end{document}